\documentclass[12pt]{article}
\usepackage{amscd,amsmath,amssymb,amstext,amsthm,exscale,latexsym}
\usepackage{graphicx}
\newcommand {\al}   {\alpha}       \newcommand {\bt}  {\beta}
\newcommand {\g }   {\gamma}       
\newcommand {\dl}   {\delta}       \newcommand {\e }  {\epsilon}
        
\newcommand {\ve}   {\varepsilon}

\newcommand {\s }   {\sigma}       
         
\newcommand {\vf }  {\varphi}

\newcommand {\pl}   {\partial}     \newcommand {\nb}  {\nabla}
\renewcommand {\det}{{\sf\,det\,}}

       \renewcommand {\lim}{{\sf\,lim\,}}
\newcommand   {\ex}{{\sf\,e}}

     \newcommand   {\diag}{{\sf\,diag\,}}

\newcommand {\MM}  {{\mathbb M}}   
   
   \newcommand {\MR}  {{\mathbb R}}
   \newcommand {\MT}  {{\mathbb T}}
\newcommand {\MU}  {{\mathbb U}}

\newcommand {\CC }  {{\cal C}}

\newcommand {\Sa}  {{\textsc{a}}}   \newcommand {\Sb}  {{\textsc{b}}}

\newcommand {\Sg}  {{\textsc{g}}}   
   
   \newcommand {\Sl}  {{\textsc{l}}}
   \newcommand {\Sn}  {{\textsc{n}}}
   
   \newcommand {\Sr}  {{\textsc{r}}}

\newtheorem{lemma}{Lemma}[section]

\newtheorem{theorem}{Theorem}[section]

\theoremstyle{definition}
\newtheorem*{cor}{Corollary}

\begin{document}
\title     {Global conformal gauge in string theory}
\author    {M. O. Katanaev
            \thanks{E-mail: katanaev@mi-ras.ru}\\ \\
            \sl Steklov Mathematical Institute,\\
           \sl ul. Gubkina, 8, Moscow, 119991, Russia}

\maketitle
\begin{abstract}
It was supposed for fifty years that the conformal gauge in string theory exists
globally on the whole string world sheet. In fact, almost all results were
obtained under this assumption. However, this statement was proved only locally
in some neighbourhood of an arbitrary point on the worldsheet, and its extension
is far from being obvious. In the present paper we prove that the conformal
gauge does exists globally on the string worldsheet represented by an infinite
strip with straight parallel boundaries.
\end{abstract}
%******************************************************************************
{\bf Introduction.}
%*******************************************************************************
The (super)string theory is one of the main fields of research in mathematical
physics for the last fifty years (see, e.g., \cite{GrScWi87,BriHen88,BarNes90}).
The basic assumption in this model is that the conformal gauge exists globally
on the whole string worldsheet represented by infinite strip with straight
boundaries. For example, this assumption provides a basis for covariant and
light cone quantization. In fact, we can say that the existence of the global
conformal gauge is crucial for the theory of strings.

After fixing the conformal gauge the (super)string becomes a consistent fruitful
and very interesting model which deserves analysis by itself. However the
question remains: is there a solution of the original covariant equations of
motion of the Nambu--Goto string which cannot be brought to the conformally flat
form? In the present paper, we prove the existence of the global conformal gauge
on the whole string worldsheet represented by infinite strip with straight
boundaries. This justifies the assumption made long ago.

The local existence of the conformal gauge is well known for a long time
(see, e.g., \cite{Petrov61,Vladim71}). It is proved by writing down equations
for transformation functions and considering their integrability conditions
which guarantee the existence of solution in some neighbourhood of an
arbitrary point. Its global extension is far from being obvious. The transition
from local to global considerations in the present paper is based on the global
existence theorem for the solution of the Cauchy problem for a two-dimensional
hyperbolic differential equations with varying coefficients (see, e.g.,
\cite{Hadama32}).
This theorem is highly nontrivial, but allows one to make global statements.

%******************************************************************************
{\bf The bosonic string.}
%*******************************************************************************
Consider two manifolds: a plane $\MR^2$ with arbitrary global
coordinates $x=(x^\al):=(x^0,x^1):=(\tau,\s)$, $\al=0,1$, and $D$-dimensional
Minkowskian space $\MR^{1,D-1}$ with Cartesian coordinates $X=(X^\Sa)$,
$\Sa=0,1,\dotsc,D-1$, $D\ge2$, and the Lorentz metric
$\eta_{\Sa\Sb}:=\diag(+-\dotsc-)$. Let there be a smooth embedding
\begin{equation}                                                  \label{ubnchs}
  X:\qquad\MR^2\supset\overline\MU\ni\qquad(\tau,\s)\mapsto
  \big(X^\Sa(\tau,\s)\big)\qquad\in\MR^{1,D-1},
\end{equation}
of some closed subset $\overline\MU$ of a plane where $\MU$ is a connected and
simply connected open subset in $\MR^2$. This embedding defines symmetric
quadratic form with components
\begin{equation}                                                  \label{enbsty}
  h_{\al\bt}:=\pl_\al X^\Sa\pl_\bt X^\Sb\eta_{\Sa\Sb}
  =\pl_\al X^\Sa\pl_\bt X_\Sa
\end{equation}
on the string worldsheet $\MM:=X(\overline\MU)$.
We assume that the embedding is such that
\begin{equation}                                                  \label{edsfwe}
\begin{split}
  (\pl_0 X)^2:=&\dot X^2:=\dot X^\Sa\dot X^\Sb\eta_{\Sa\Sb}>0,
\\
  (\pl_1 X)^2:=&X^{\prime2}:=X^{\prime\Sa}X^{\prime\Sb}\eta_{\Sa\Sb}<0,
\end{split}
\end{equation}
where the dot and prime denote differentiations with respect to $\tau$ and $\s$,
respectively, on $\MU$. Here and in
what follows indices $\Sa,\Sb,\dotsc$ are often omitted. So global coordinates
$\tau,\s$ on $\MU$ are timelike and spacelike, respectively. Then the
determinant of the induced quadratic form is negative
\begin{equation}                                                  \label{eqjfyh}
  h:=\det h_{\al\bt}=\dot X^2X^{\prime2}-(\dot X,X')^2<0,
\end{equation}
where parenthesis denote the usual scalar product in $\MR^{1,D-1}$. Now the
embedding (\ref{ubnchs}) defines the Lorentzian metric on the string worldsheet
interior $\MU$ with signature $(+-)$.

Open string is the embedding (\ref{ubnchs}) of the closed straight strip
\begin{equation}                                                  \label{bdvgtr}
  -\infty<\tau<\infty,\qquad 0\le\s\le\pi
\end{equation}
with properties (\ref{edsfwe}). This strip is vertical if $\tau$ and $\s$
coordinate axes are depicted by vertical and horizontal straight lines on a
plain $\MR^2$, respectively.

Closed string is the embedding (\ref{ubnchs}) of the closed straight vertical
strip (fundamental domain)
\begin{equation}                                                  \label{qndbfu}
  -\infty<\tau<\infty,\qquad -\pi\le\s\le\pi
\end{equation}
with identified boundaries. There are many ways to identify smoothly the
boundaries (\ref{qndbfu}). In string theory, we, first, impose the conformal
gauge on the metric on the same strip (\ref{qndbfu}) and, second, impose the
smooth periodicity conditions
\begin{equation}                                                  \label{annfgh}
  \pl^k_1 X^\Sa\big|_{\s=-\pi}=\pl^k_1 X^\Sa\big|_{\s=\pi},\qquad\forall\Sa,
  \forall\tau,\quad k=0,1,2,\dotsc,
\end{equation}
up to the needed order. It is the prime aim of the present paper to prove that
the conformal gauge on the same strips does exist.

If infinite strips in the $\tau,\s$ plane have curved boundaries, then all of
them are diffeo\-morphic to strips (\ref{bdvgtr}) or (\ref{qndbfu}). Thus we did
not loose generality by specifying the coordinate range on the $\tau,\s$ plane.

The dynamics of the Nambu--Goto string is governed by the action proportional
to the string worldsheet area
\begin{equation}                                                  \label{ubxvgy}
  S_{\Sn\Sg}:=-\int_{\overline\MU}\!\!dx\sqrt{|h|}
  =-\int_{\overline\MU}\!\!d\tau d\s\sqrt{\displaystyle(\dot X,X')^2
  -\dot X^2X^{\prime2}}.
\end{equation}
It implies the Euler--Lagrange equations
\begin{equation}                                                  \label{uvbsju}
  \frac1{\sqrt{|h|}}\frac{\dl S_{\Sn\Sg}}{\dl X_\Sa}=
  \square_{(h)} X^\Sa=h^{\al\bt}\nb_\al\nb_\bt X^\Sa
  =\frac1{\sqrt{|h|}}\pl_\al\left(\sqrt{|h|}h^{\al\bt}\pl_\bt X^\Sa\right)=0,
\end{equation}
where the two-dimensional wave operator $\square_{(h)}$ is build
by the induced metric $h_{\al\bt}$ (\ref{enbsty}) and $\nb_\al$ is the
covariant derivative with respective Christoffel's symbols.

We assume that ends of an open string are free, and then the action
(\ref{ubxvgy}) implies also the boundary conditions
\begin{equation}                                                  \label{ubvbxg}
  s^\bt\pl_\bt X^\Sa\big|_{\s=0,\pi}=0,
\end{equation}
where $s^\al$ are components of the spacelike vector which is perpendicular
to the boundaries with respect to the induced metric.
The action (\ref{ubxvgy}) does not yield any boundary condition for a closed
string. Instead, we have periodicity conditions (\ref{annfgh}) imposed by hands.

In string theory, the crucial role is played by the possibility to impose
global conformal gauge
\begin{equation}                                                  \label{edbfht}
  h_{\al\bt}=\ex^{2\phi}\eta_{\al\bt},\qquad\eta_{\al\bt}:=\diag(+-),
\end{equation}
where $\phi(x)$ is some sufficiently smooth function, on the whole string
worldsheet. The aim of the present paper is to prove that this conformal gauge
can be imposed on the same strips (\ref{bdvgtr}) and (\ref{qndbfu}) both for
open and closed strings with the same straight boundaries.

%******************************************************************************
{\bf The idea of the proof.}
%*******************************************************************************
We construct two orthogonal vector fields: the timelike $t=t^\al\pl_\al$ and
spacelike $s=s^\al\pl_\al$ with properties
\begin{equation}                                                  \label{axncjh}
  (t,s)=0,\qquad t^2+s^2=0,\qquad t^2>0,\qquad\forall x\in\MU,
\end{equation}
where
\begin{equation*}
  (t,s):=t^\al s^\bt h_{\al\bt},\qquad t^2:=(t,t),\qquad s^2:=(s,s).
\end{equation*}
Then we find conditions for commutativity of these vector fields: $[t,s]=0$.
The next step is to find two families of integral curves
$x^\al(\tilde\tau,\tilde\s)$ defined by the system of equations
\begin{equation}                                                  \label{ancjdy}
  \frac{\pl x^\al}{\pl\tilde\tau}=t^\al,\qquad
  \frac{\pl x^\al}{\pl\tilde\s}=s^\al,
\end{equation}
where $\tilde\tau$ and $\tilde\s$ are parameters along integral curves of vector
fields $t$ and $s$, respectively. The integrability conditions for this system
are fulfilled on the whole $\MU$:
\begin{equation*}
  \frac{\pl^2 x^\al}{\pl\tilde\tau\pl\tilde\s}
  -\frac{\pl^2 x^\al}{\pl\tilde\s\pl\tilde\tau}=\frac{\pl s^\al}{\pl\tilde\tau}
  -\frac{\pl t^\al}{\pl\tilde\s}=t^\bt\pl_\bt s^\al-s^\bt\pl_\bt t^\al
  =[t,s]^\al=0,
\end{equation*}
due to the commutativity of vector fields. Consequently, there is a
nondegenerate coordinate transformation $(\tau,\s)\mapsto(\tilde\tau,\tilde\s)$
on the whole worldsheet $\MU$.

In the new coordinate system, the induced metric $\tilde h_{\al\bt}$ is
conformally flat due to Eqs.(\ref{axncjh}):
\begin{equation}                                                  \label{ehsdhg}
\begin{split}
  \tilde h_{00}=&h_{\al\bt}\frac{\pl x^\al}{\pl\tilde\tau}
  \frac{\pl x^\bt}{\pl\tilde\tau}=t^2,
\\
  \tilde h_{01}=&h_{\al\bt}\frac{\pl x^\al}{\pl\tilde\tau}
  \frac{\pl x^\bt}{\pl\tilde\s}=(t,s)=0,
\\
  \tilde h_{11}=&h_{\al\bt}\frac{\pl x^\al}{\pl\tilde\s}
  \frac{\pl x^\bt}{\pl\tilde\s}=s^2=-t^2.
\end{split}
\end{equation}
The final step is the analysis of domains of the definition of parameters
$\tilde\tau$ and $\tilde\s$ of integral curves (\ref{ancjdy}) which are new
coordinates.

%******************************************************************************
{\bf Infinite string.}
%*******************************************************************************
First, we consider embedding (\ref{ubnchs}) where $\MU=\MR^2$. Arbitrary
timelike and spacelike tangent vectors $T$ and $S$ to the string worldsheet in
the embedding space $\MR^{1,D-1}$ can be decomposed on $\dot X$ and $X'$:
\begin{equation}                                                  \label{abnxhh}
\begin{split}
  T=&A(\cosh\vf\dot X+\sinh\vf X'),
\\
  S=&B(\sinh\psi\dot X+\cosh\psi X'),
\end{split}
\end{equation}
where $A(x)>0$, $B(x)>0$ and $\vf(x),\psi(x)\in\MR$ are some functions.
\begin{lemma}                                                     \label{lsjdhg}
Vector fields $T$ and $S$ on $\MU$ satisfy equalities
\begin{equation}                                                  \label{abcnft}
  (T,S)=0,\qquad T^2+S^2=0,
\end{equation}
if and only if vector field $S$ is given by Eq.(\ref{abnxhh}) with arbitrary
functions $B>0$ and $\psi\in\MR$, and the second vector field has the form
\begin{equation}                                                  \label{anbsgt}
  T=-\frac B{\sqrt{|h|}}\big[\cosh\psi X^{\prime2}+\sinh\psi(\dot X,X')\big]\dot X
  +\frac B{\sqrt{|h|}}\big[\sinh\psi\dot X^2+\cosh\psi(\dot X,X')\big]X'.
\end{equation}
\end{lemma}
The proof is given in \cite{Katana19B}.

The differential map of the embedding $\MU\hookrightarrow\MR^{1,D-1}$ acts on
vectors as follows
\begin{equation*}
  \MT(\MU)\ni\quad t=t^\al\pl_\al,~s=s^\al\pl_\al~\mapsto~
  T=t^\al\pl_\al X^\Sa\pl_\Sa,~S=s^\al\pl_\al X^\Sa\pl_\Sa\quad
  \in\MT(\MR^{1,D-1}),
\end{equation*}
where $t$ and $s$ are vector fields on $\MU$. Comparing the above
formulae with Eqs.(\ref{abnxhh}) and (\ref{anbsgt}) allows us to define
vector fields on $\MU$:
\begin{equation}                                                  \label{abcndg}
\begin{split}
  t=&-\frac B{\sqrt{|h|}}\big[\cosh\psi X^{\prime2}
  +\sinh\psi(\dot X,X')\big]\pl_0
  +\frac B{\sqrt{|h|}}\big[\sinh\psi\dot X^2+\cosh\psi(\dot X,X')\big]\pl_1.
\\
  s=&~~B\sinh\psi\pl_0+B\cosh\psi\pl_1.
\end{split}
\end{equation}
These vectors can be easily rewritten in the form
\begin{equation}                                                  \label{avsfrd}
  t=\ve^{\al\bt}s_\bt\pl_\al,\qquad s=s^\al\pl_\al,
\end{equation}
where $\ve^{\al\bt}$ is the totally antisymmetric second rank tensor,
$\ve^{01}=-1/\sqrt{|h|}$, and components $s^\al$ are arbitrary.
\begin{lemma}                                                     \label{lkwioi}
Vector fields $t$ and $s$ on $\MU$ related by equalities (\ref{avsfrd}) commute
if and only if
\begin{equation}                                                  \label{amvnfu}
  t_\al=\frac{\pl_\al\chi}{\pl\chi^2},\qquad
  \pl\chi^2:=h^{\al\bt}\pl_\al\chi\pl_\bt\chi>0,
\end{equation}
where $\chi$ is a nontrivial solution of the wave equation
\begin{equation}                                                  \label{amnfht}
  \Box_{(h)}\chi:=h^{\al\bt}\nb_\al\nb_\bt\chi=0,
\end{equation}
satisfying $\pl\chi^2>0$.

For any nontrivial solution of Eq.(\ref{amnfht}) satisfying the condition
$\pl\chi^2>0$, vector fields (\ref{avsfrd}) and (\ref{amvnfu}) commute.
\end{lemma}
The proof is given in \cite{Katana19B}.

Thus commuting vector fields $t$ and $s$ with properties (\ref{axncjh}) have
generally the following form
\begin{equation}                                                  \label{andytr}
  t=\frac{h^{\al\bt}\pl_\bt\chi}{\pl\chi^2}\pl_\al,\qquad
  s=\frac{\ve^{\al\bt}\pl_\bt\chi}{\pl\chi^2}\pl_\al,
\end{equation}
where $\chi$ is an arbitrary solution of the wave equation (\ref{amnfht}) such
that $\pl\chi^2>0$.

Suppose that the determinant of the induced metric $h_{\al\bt}$ in nonzero on
the whole plane $(\tau,\s)\in\MR^2$ and separated from $0$ and $\pm\infty$ at
infinity:
\begin{equation}                                                  \label{abnshg}
  0<\e\le\underset{\tau^2+\s^2\to\infty}\lim|\det h_{\al\bt}|\le M<\infty,
\end{equation}
where $\e$ and $M$ are some constants. It is well known that the Cauchy problem
for the hyperbolic equation (\ref{amnfht}) has unique solution $\chi$ on the
whole plain, if the Cauchy data are given on a spacelike curve, say, $\tau=0$
(see, e.g., \cite{Hadama32}, book IV, ch.\ I).
It is easily verified that there exist such Cauchy data that the inequality
$\pl\chi^2>0$ holds everywhere. This implies that nontrivial solution of the
wave equation  (\ref{amnfht}) exists on the whole plane $\MR^2$. There are many
such solutions, and they are parameterized by the Cauchy data.

Thus the vector fields $s$ and $t$ are given on the whole plane $\MR^2$. The
inequality $t^2>0$ (\ref{axncjh}) implies that component $t^0$ is separated from
zero and bounded on the plane including infinity. Thus Eqs.(\ref{ancjdy}) imply
\begin{equation*}
  \frac{\pl\tau}{\pl{\tilde\tau}}=t^0\qquad\Rightarrow\qquad
  \tilde\tau\sim\int^\infty\frac{d\tau}{t^0}.
\end{equation*}
The last integral is divergent and thus the coordinate $\tilde\tau$ runs over
the whole real line $\MR$. Similar statement is valid for the space coordinate
$\tilde\s$. Consequently, new coordinates cover the whole plane
$(\tilde\tau,\tilde\s)\in\MR^2$.

Moreover, it is known that any surface with a Lorentzian metric can be
globally isometrically embedded in flat Minkowskian space $\MR^{1,D-1}$ of
sufficiently large dimension. Since the dimension $D$ in Lemma \ref{lsjdhg} is
not fixed, we obtain
\begin{theorem}                                                   \label{tdgefr}
Let an arbitrary metric $h_{\al\bt}$ of Lorentzian signature be given on the
whole plane $\MR^2$. Let it be nondegenerate at infinity (\ref{abnshg}). Then
there exists a surjective diffeomorphism on the plane
\begin{equation}                                                  \label{ehgdrv}
  \MR^2\ni\quad (x^\al)\mapsto\big(\tilde x^\al(x)\big)\quad\in\MR^2
\end{equation}
such that metric $h_{\al\bt}$ in new coordinate system has conformally flat form
\begin{equation}                                                  \label{ehhhgd}
  \tilde h_{\al\bt}:=h_{\g\dl}\frac{\pl x^\g}{\pl\tilde x^\al}
  \frac{\pl x^\dl}{\pl\tilde x^\bt}=\ex^{2\phi}\eta_{\al\bt},
\end{equation}
where $\phi(\tilde x)$ is some function on $\MR^2$ separated from $\pm\infty$
at infinity $\tilde\tau^2+\tilde\s^2\to\infty$.
\end{theorem}

\begin{cor}
Let
\begin{equation}                                                  \label{anvbft}
  \widetilde\MU_0:=\lbrace (\tilde\tau,\tilde\s)\in\MR^2:\quad
  \tilde\s\in[\tilde\s_1,\tilde\s_2],~\tilde\tau\in\MR\rbrace
\end{equation}
be closed vertical strip with straight boundaries on the plane of new
coordinates $\tilde\tau,\tilde\s$ and assumptions of theorem \ref{tdgefr} hold.
Then there exists diffeomorphism (\ref{ehgdrv}) of a closed domain
$(\tau,\s)\in\overline\MU\subset\MR^2$ bounded by integral curves
$x(\tilde\tau,\tilde\s_{1,2})$:
\begin{equation*}
  \frac{\pl x(\tilde\tau,\tilde\s_{1,2})}{\pl\tilde\tau}=t_{1,2},
\end{equation*}
where $t_{1,2}$ are inverse images of vector fields $\pl/\pl\tilde\tau$ to the
boundaries of $\widetilde\MU_0$.
\qed\end{cor}

To clarify the arbitrariness in coordinates $\tilde\tau$, $\tilde\s$ defined by
the function $\chi$ we consider

{\bf Example}. Let the induced metric be conformally flat:
\begin{equation}                                                  \label{abdnft}
  h_{\al\bt}dx^\al dx^\bt=\ex^{2\phi}(d\tau^2-d\s^2)
  =\ex^{2\phi}d\xi d\eta,
\end{equation}
where light cone coordinates $\xi:=\tau+\s$, $\eta:=\tau-\s$ are introduced.
Then the wave equation (\ref{amnfht}) is reduced to the flat d'Alembert equation
$(\pl^2_0-\pl^2_1)\chi=0$.
Its general solution is given by two arbitrary sufficiently smooth functions
\begin{equation*}
  \chi=F(\xi)+G(\eta).
\end{equation*}
We choose only the functions which satisfy inequality
\begin{equation*}
  \pl\chi^2=4\ex^{-2\phi}F'G'>0\qquad\Rightarrow\qquad F'G'>0,
\end{equation*}
where prime denotes differentiation by the corresponding argument. Then the
metric is
\begin{equation*}
  \tilde h_{\al\bt}d\tilde x^\al d\tilde x^\bt
  =\frac{\ex^{2\phi}}{4F'G'}d\tilde\xi d\tilde\eta.
\end{equation*}
It corresponds to the conformal transformation
\begin{equation*}
  \tilde\xi:=2F(\xi),\qquad\tilde\eta:=2G(\eta).
\end{equation*}
We see that the arbitrariness in definition of the vector fields described in
Lemmas \ref{lsjdhg} and \ref{lkwioi} corresponds to conformal transformations
on the string worldsheet.
\qed

Thus to find the diffeomorphism (\ref{ehgdrv}) in explicit form for a given
metric $h_{\al\bt}$, we have to (i) find a nontrivial solution of the wave
equation (\ref{amnfht}), (ii) construct the vector fields $t$ and $s$ using
Eqs.(\ref{amvnfu}), (\ref{avsfrd}), and (iii) find a general solution of the
system of equations (\ref{ancjdy}). We have proved that this problem does have
many solutions (the whole arbitrariness is contained in the choice of nontrivial
solution of the wave equation).

%******************************************************************************
{\bf Open string.}
%*******************************************************************************
Now we consider an open string which worldsheet $\overline\MU$ is an infinite
strip on the plane $(\tau,\s)\in\MR^2$ with two, probably, curved left $\g_\Sl$
and right $\g_\Sr$ boundaries. The induced metric on the
boundaries is degenerate, and results of the previous section must be revised.
First, we assume that metric is not degenerate and return to this problem later.

If the metric is nondegenerate on $\overline\MU$ including the boundaries, then
we continue it on the whole plain in some sufficiently smooth manner. As the
consequence of theorem \ref{tdgefr} there is a diffeomorphism (\ref{ehgdrv})
after which the metric becomes conformally flat. The problem is that the
boundaries $\g_{\Sl,\Sr}$ on the plain $\tilde\tau,\tilde\s$ may be not
straight vertical lines. However there are residual diffeomorphisms in the
form of conformal maps of $\tilde\tau,\tilde\s$ coordinates. We now show that it
is enough to straighten the strip.

Let boundary equations after diffeomorphism
$(\tau,\s)\mapsto(\tilde\tau,\tilde\s)$ be
 \begin{equation}                                                 \label{eaffqj}
   \g_\Sl:\quad\tilde\eta=\tilde\eta_\Sl(\tilde\xi),\qquad
   \g_\Sr:\quad\tilde\eta=\tilde\eta_\Sr(\tilde\xi),\qquad
   \tilde\xi\in\MR,
\end{equation}
where functions $\tilde\eta_{\Sl,\Sr}\in\CC^1(\MR)$ have properties:
\begin{equation*}
  \tilde\eta_\Sl>\tilde\eta_\Sr,\qquad
  0<\e\le\frac{d\tilde\eta_{\Sl,\Sr}}{d\tilde\xi}\le M<\infty,\qquad\e,M\in\MR
\end{equation*}
for all $\tilde\xi\in\MR$ including infinite points.
\begin{theorem}                                                   \label{thhdgy}
The conformal transformation
\begin{equation*}
  \hat\xi=F(\tilde\xi),\qquad\hat\eta=G(\tilde\eta),\qquad F,G\in\CC^1(\MR),
\end{equation*}
such that the boundaries (\ref{eaffqj}) of an open string worldsheet become
straight vertical lines
\begin{equation}                                                  \label{ajhdgt}
   \g_\Sl:\quad\hat\eta=\hat\xi,\qquad
   \g_\Sr:\quad\hat\eta=\hat\xi-2\pi,\qquad\hat\xi\in\MR
\end{equation}
on the plain $\hat\xi,\hat\eta\in\MR^2$ exists.
\end{theorem}
The proof is given in \cite{Katana19B}.

Now we discuss an open Nambu--Goto string for which the induced metric on the
boundaries is degenerate due to boundary conditions. Let us parameterize metric
$h_{\al\bt}$ by its determinant $\varrho^4$ and ``unimodular metric''
$k_{\al\bt}$:
\begin{equation}                                                  \label{abcndf}
  h_{\al\bt}:=\varrho^2k_{\al\bt},\qquad \det k_{\al\bt}:=-1.
\end{equation}
It implies that the variable $\varrho$ is the scalar density of degree $-1/2$
and unimodular metric is the second rank tensor density of degree 1.

The boundary condition (\ref{ubvbxg}) has the form
\begin{equation*}
  n^0\dot X^\Sa+n^1X^{\prime\Sa}=0\qquad\Rightarrow\qquad
  X^{\prime\Sa}=-\frac{n^0}{n^1}\dot X^\Sa,
\end{equation*}
because the normal vector $(n^0,n^1)$ must be spacelike and consequently
$n^1\ne0$. As the consequence, the metric degenerates on the boundaries:
\begin{equation*}
  \det h_{\al\bt}=\rho^4=\dot X^2 X^{\prime2}-(\dot X,X')^2\to0.
\end{equation*}
Therefore vector fields $t$ and $s$ (\ref{axncjh}) become null
\begin{equation*}
  t^2=\rho^2k_{\al\bt}t^\al t^\bt\to0,\qquad
  s^2=\rho^2k_{\al\bt}s^\al s^\bt\to0
\end{equation*}
at the ends of the string because $\rho\to0$.

Now we construct new vector fields for the unimodular metric $k_{\al\bt}$
satisfying relations
\begin{equation}                                                  \label{avsbsf}
  (t,s):=k_{\al\bt}t^\al s^\bt=0,\qquad t^2+s^2=0,
\end{equation}
where $t^2:=k_{\al\bt}t^\al t^\bt$ and $s^2:=k_{\al\bt}s^\al s^\bt$.

Equalities (\ref{avsbsf}) are equivalent to original equations (\ref{axncjh}) in
internal points of $\MU$ and extended on boundaries $\pl\MU$ by continuity.
Now we prove that new vector fields $t$ and $s$ exist and coincide with the
original ones for $h_{\al\bt}$.

Formulae (\ref{ehsdhg}) for metric components after the coordinate
transformation have the same form. In addition,
\begin{equation*}
  \tilde h_{00}=-\tilde h_{11}=\varrho^2k_{\al\bt}t^\al t^\bt,\qquad
  \tilde h_{01}=0.
\end{equation*}

Function $\chi$ in new variables must satisfy the equation which does not
depend on $\rho$ \cite{Katana19B}:
\begin{equation}                                                  \label{ennngh}
  \square_{(k)}\chi:=\pl_\al\big(k^{\al\bt}\pl_\bt\chi\big)=0.
\end{equation}
This wave equation has many solutions on the whole plain $\MR^2$ because depends
on nondegenerate unimodular metric $k_{\al\bt}$. It implies that vector fields
$t$ and $s$ exist and do not depend on $\varrho$:
\begin{equation}                                                  \label{eersew}
  t^\al=\frac{k^{\al\bt}\pl_\bt\chi}{k^{\g\dl}\pl_\g\chi\pl_\dl\chi}\qquad
  s^\al=\frac{\hat\ve^{\al\bt}\pl_\bt\chi}{k^{\g\dl}\pl_\g\chi\pl_\dl\chi},
\end{equation}
where
\begin{equation*}
  \hat\ve^{\al\bt}=\rho^2\ve^{\al\bt}=
  \begin{pmatrix} 0 & -1 \\ 1 & ~~0 \end{pmatrix}
\end{equation*}
is the totally antisymmetric tensor density of degree $-1$.

One can easily verify that in new variables the Euler--Lagrange equations for
bosonic string (\ref{uvbsju}) do not depend on $\varrho$
\begin{equation}                                                  \label{avdbcf}
  \sqrt{|h|}\square_{(h)} X^\Sa=\pl_\al\big(\sqrt{|h|}h^{\al\bt}
  \pl_\bt X^\Sa\big)=\pl_\al\big(k^{\al\bt}\pl_\bt X^\Sa\big)=0.
\end{equation}
They must be solved with the boundary condition
\begin{equation}                                                  \label{avdbck}
  n^\al\pl_\al X^\Sa\big|_{\g_{\Sl,\Sr}}=0,
\end{equation}
which does not depend on $\rho$ too.

Consequently, the problem is reduced to solution of Eqs.(\ref{avdbcf}) with
boundary conditions (\ref{avdbck}) for an open Nambu--Goto string. To make the
transformation of coordinates $(\tau,\s)\mapsto(\tilde\tau,\tilde\s)$ we have to
find the unimodular metric $k_{\al\bt}$ for a given metric $h_{\al\bt}$, choose
a solution $\chi$ of the wave equation (\ref{ennngh}) satisfying the condition
$k^{\g\dl}\pl_\g\chi\pl_\dl\chi>0$, construct the vector fields $t$ and $s$
using formulae (\ref{eersew}), and, finally, integrate Eqs.(\ref{ancjdy}).
Therefore the corollary of theorem \ref{tdgefr} is valid also for an open
string. If needed, after solution of this problem, one can compute the
conformal factor for the induced metric (\ref{ehhhgd}) which goes to zero at the
boundaries. Thus we proved the existence of the global conformal gauge for an
open string.

%******************************************************************************
{\bf Closed string.}
%*******************************************************************************
In the initial coordinates $\tau,\s\in\MR^2$, the worldsheet of a closed string
is given by an infinite strip with timelike boundaries which are identified. The
identification can be performed in many ways and therefore requires definition.
Here we describe the method adopted in string theory.

We showed in the previous section that there is the global diffeomorphism
$(\tau,\s)\mapsto(\hat\tau,\hat\s)$ which maps an arbitrary infinite strip with
timelike boundaries on the vertical strip with straight boundaries where metric
takes the conformally flat form. The same procedure can be performed for the
fundamental domain of a closed string. Without loss of generality, we assume
that boundaries go through points $\hat\s=\pm\pi$. Then the boundary
identification is written as the periodicity condition:
\begin{equation}                                                  \label{abbcvf}
  \left.\frac{\pl^k X^\Sa}{\pl\hat\s^k}\right|_{\hat\s=-\pi}=
  \left.\frac{\pl^k X^\Sa}{\pl\hat\s^k}\right|_{\hat\s=\pi},\qquad
  \forall\Sa,\quad\forall\hat\tau,\quad k=0,1,2,\dotsc.
\end{equation}
In the initial coordinate system this condition is written in the covariant form
\begin{equation}                                                  \label{ancbgf}
  \nb_s^k X^\Sa\big|_{\hat\s=-\pi}=\nb_s^k X^\Sa\big|_{\hat\s=\pi},
\end{equation}
where $\nb_s:=s^\al\nb_\al$ is the covariant derivative for the Levi--Civita
connection along the normal vector field $s$ which is the pullback of the vector
field $\pl/\pl\hat\s$ under the diffeomorphism
$(\tau,\s)\mapsto(\hat\tau,\hat\s)$.
However it is not clear at the beginning for which value of $\tau$ on the left
and right the identification takes place, because we have to find the
diffeomorphism $(\tau,\s)\mapsto(\hat\tau,\hat\s)$ explicitly. Simply speaking
we firstly transform the metric to conformally flat form and afterwards perform
the natural gluing.

%******************************************************************************
{\bf Conclusion.}
%*******************************************************************************
It was assumed for many years that there exists the global conformal gauge in
string theory though this statement was proved only locally. We proved the
global existence of the conformal gauge for infinite, open, and closed strings.
As a byproduct, we proved global existence of the conformal gauge for a general
two-dimensional Lorentzian metric which is not necessarily induced by an
embedding.

%\bibliography{3dgrav,book,gravity,math,my,qft}
%\bibliographystyle{unsrt}
\end{document}